# Free and forced wave propagation in beam lattice metamaterials with viscoelastic resonators

F Vadalà[1], A Bacigalupo[2], M Lepidi[1] and L Gambarotta[1]

[1]DICCA, University of Genoa, Via Montallegro 1, 16145, Genova, Italy,

[2]IMT School for Advanced Studies, Piazza S. Francesco 19, 55100, Lucca, Italy.

E-mail:  francesca.vadala@edu.unige.it, andrea.bacigalupo@imtlucca.it,

marco.lepidi@unige.it, luigi.gambarotta@unige.it.

**Abstract.** Beam lattice materials are characterized by a periodic microstructure realizing a geometrically regular pattern of elementary cells. In these microstructured materials, the dispersion properties governing the free dynamic propagation of elastic waves can be studied by formulating parametric lagrangian models and applying the Floquet-Bloch theory. Within this framework, governing the wave propagation by means of spectral design techniques and/or energy dissipation mechanisms is a major issue of theoretical and applied interest. Specifically, the wave propagation can be inhibited by purposely designing the microstructural parameters to open band gaps in the material spectrum at target center frequencies. Based on these motivations, a general dynamic formulation for determining the dispersion properties of beam lattice metamaterials, equipped with local resonators is presented. The mechanism of local resonance is realized by tuning periodic auxiliary masses, viscoelastically coupled with the beam lattice microstructure. As peculiar aspect, the viscoelastic coupling is derived by a mechanical formulation based on the Boltzmann superposition integral, whose kernel is approximated by a Prony series. Consequently, the free propagation of damped waves is governed by a linear homogeneous system of integral-differential equations of motion. Therefore, differential equations of motion with frequency-dependent coefficients are obtained by applying the bilateral Laplace transform. The corresponding complex-valued branches characterizing the dispersion spectrum are determined and parametrically analyzed. Particularly, the spectra corresponding to Taylor series approximations of the equation coefficients are investigated. The standard dynamic equations with linear viscous damping are recovered at the first-order approximation. Increasing approximation orders determine non-negligible spectral effects, including the occurrence of pure-damping spectral branches. Finally, the forced response to harmonic mono-frequent external forces is investigated. The metamaterial responses to non-resonant, resonant and quasi-resonant external forces are compared and discussed from a qualitative and quantitative viewpoint.

---

Corresponding author: Andrea Bacigalupo, mail: andrea.bacigalupo@imtlucca.it

## 1. Introduction

Microstructured material and metamaterial science is a challenging frontier of theoretical and applied research that is currently attracting growing interest by the scientific community of solid and structural mechanicians [1],[2],[3],[4]. Specifically, the conceptualization and development of novel materials, characterized by smart or unconventional functionalities, are continuously propelled by the recent extraordinary developments in the technological fields of super-computing, micro-engineering and high-precision manufacturing [5],[6],[7]. As valuable result of this successful research trend, a new generation of architected composites is deeply transforming and rapidly remodeling the traditional paradigms of rational design in a variety of technical multidisciplinary applications across all the classical and advanced branches of mechanics, including – among the others – extreme mechanics, nanomechanics, mechatronics, acoustics, thermomechanics, biomechanics [8],[9],[10],[11].

Within this stimulating framework, composite materials and metamaterials with periodic cellular microstructure are being purposely designed to achieve superior effective properties, outperforming the elasto-dynamic characteristics of the ingredient materials building each microstructured cell [12],[13],[14]. Significant achievements have been obtained in tailoring unusual properties or exotic performances, including for instance super lightness-to-strength ratios, strong auxeticity, synclastic bending curvatures, giant hysteresis, morphing and multi-stability, negative indexes of dynamic refraction, non-reciprocal vibration propagation, broadband sound absorption, controllable wave guiding, obstacle cloaking, low-frequency noise filtering, energy focusing or harvesting [15],[16],[17],[18],[19],[20],[21],[22],[23],[24].

Focusing on the dynamic response of periodic cellular media, a major issue of mechanical interest consists in governing the wave propagation by means of spectral design techniques and/or energy dissipation mechanisms. To this purpose, the wave propagation around certain target center-frequencies can be inhibited by finely designing the microstructural parameters in order to open band gaps in the material dispersion spectrum. Basically, the design problem can be stated either as an inverse problem or as an optimization problem. According to the former approach, the linear eigenproblem governing the free wave propagation can be stated, solved and – in principle – inverted to analytically assess the microstructural parameters satisfying desired spectral requirements, like the existence, position and amplitude of a given harmonic component or certain pass or stop bands in the frequency spectrum [25],[26],[27]. Of course, solving the inverse problem may be not straightforward and – in the general case – some forms of mathematical approximations (e.g. asymptotic perturbation-based solutions) must be accepted to preserve the analytical assessment of the design variables [28],[29]. According to the latter approach, a suited objective or multi-objective function can be formulated, so that its maximization or minimization allows the numerical identification of the optimal solution in the multidimensional space of the design parameters [31],[32],[33]. Generally, a proper mathematical surrogation of the objective function may help in reducing the computational costs and accelerating the algorithmic convergence [34]. Independently of the particular approach, the existence, uniqueness and admissibility of the design solution must be discussed. In the spectral design of cellular periodic materials for low-frequency wave filtering, the extent and dimensionality of the optimal solution domain can be significantly enlarged by designing *acoustic metamaterials*, realized by introducing auxiliary massive oscillators, mechanically coupled to the cell microstructure. Indeed, if the oscillators are properly tuned (*local resonators*), their dynamic interaction with the microstructure ends up opening a band gap in the dispersion spectrum [35],[36]. It can be demonstrated that the achievable band gap is nearly centered at the oscillator frequency, with a bandwidth almost directly proportional to its mass [29],[37],[38]. Accordingly, different parametric forms of spectral design tend to realize the desirable optimum of maximizing the band gap amplitude centered at the lowest center frequency by introducing heavy oscillators, weakly coupled with the microstructure [37],[39]. It may be worth remarking that the peculiar mechanism of local resonance can be virtuously exploited to achieve other desirable effects, including modal localization, transmission amplification, image lensing, wave trapping and edging [40],[41].

Starting from this scientific background, formulating microstructural models of locally-resonant acoustic metamaterials is an active research field, whose development is motivated by some open investigation issues. First, a general improvement in the elastodynamic description of the linear and nonlinear dissipation mechanisms occurring in infinite periodic phononic systems has been recognized as the theoretical key point for the future advances in the energetically consistent modelization and spectral design of acoustic metamaterials [14]. Second, a completely new class of mechanical meta-behaviours has been postulated to be developable in the next few years, by exploiting the virtuous contrast and synergy among constituent ingredient materials featured by strongly dissimilar elastic, plastic and viscous properties [13]. Based on these motivations, the paper presents a beam lattice formulation for describing the wave dynamics of an acoustic metamaterial, originated by a periodic non dissipative microstructure viscoelastically coupled with local resonators. As original aspect, the viscoelastic coupling is consistently derived by a physical-mathematical construct based on the Boltzmann superposition integral, whose kernel is approximated by a Prony series [42]. Consequently, the non-conservative wave propagation is governed by a linear homogeneous system of integral-differential equations of motion. This integral description of the viscoelastic metamaterial dissipation enriches the classic formulations of viscous damping, sometimes following the rheological Rayleigh or Maxwell models [43],[44], which can be recovered for particular parameter values and low-order approximations of the governing equations in the transformed Laplace space. According to the so-called *inverse method* [14], the associated non-polynomial eigenproblem is solved in the space of complex frequencies and real wavevectors. Therefore, the complex frequencies can be expressed in terms of real-valued damped frequencies and damping ratio [45],[46]. This solution approach differs from the so-called *direct method* [14], in which the governing eigenproblem is solved in the space of real frequencies and complex wavevectors [47],[48],[49],[50],[51]. Specifically, the latter approach has been successfully employed in studying lossy phononic crystals [48] and periodic heterogeneous materials [49] containing viscoelastic phases. Moreover, the viscoelastic effects on the wave dispersion properties have been theoretically and experimentally investigated in acoustic metamaterials, by introducing the material losses through stress–strain relations in terms of hereditary or Duhamel integrals [50],[51],[52].

In order to characterize the free and forced propagation of damped waves in the acoustic metamaterials, a Lagrangian linear model of the periodic beam lattice microstructure, visco-elastically coupled with local resonators, is formulated (Section 2). Therefore, the dynamic problem concerning the wave propagation of damped waves is stated according to the Floquet-Bloch theory (Section 3). First, the complex dispersion spectrum characterizing the free dynamics is determined for the beam lattice with quadrilateral elementary cell, and the effects of different approximations of the coupling relaxation functions are parametrically analysed, with reference to the exact dispersion curves (Section 4). Second, the forced response to harmonic mono-frequent external sources is investigated in the frequency and time domain for the fundamental cases of non-resonant, resonant and quasi-resonant external forces (Section 5). Finally, concluding remarks are pointed out.

## 2. Governing equations of the beam lattice model

The periodic metamaterials with viscoelastic resonators can be based on different planar topologies described by quadrilateral or triangular beam lattices (Figure 1). The periodic cell of the metamaterial, with characteristic size $a$ and unitary out-of-plane depth $d$, is featured by a centrosymmetric microstructure realized by a massive rigid ring, with radius $R$, mass $M_1$ and rotational inertia $J_1$. Each ring is connected with the rings of adjacent cells by $n$ identical flexible and light ligaments of length $l$, width $w$ and Young's modulus $E_s$. The ring-ligament connection is supposed to realize a perfectly rigid joint. Each ring hosts a heavy disk with radius $r$, co-centered with the ring center and embedded in a soft viscoelastic annular filler. This circular massive inclusion plays the role of local resonator. The mass and rotational inertia of the local resonator are $M_2$ and $J_2$, respectively. The motion of the rigid ring is described by the in-plane displacement vector $\mathbf{u}$ and the rotation $\phi$, referred to the internal point located at the ring barycenter. The motion of the resonator is described by the in-plane displacement vector $\mathbf{v}$ and the rotation $\theta$. The integral-differential equations governing the forced response of motion of a reference cell read

$$M_1\ddot{\mathbf{u}} + \int_{-\infty}^{t} k_d(t-\tau)\frac{d}{d\tau}(\mathbf{u}-\mathbf{v})\,d\tau + \sum_{i=1}^{n}\left[\mathbf{K}_i\left(\mathbf{u}-\mathbf{u}_i\right) + \mathbf{k}_i\left(\phi+\phi_i\right)\right] = \mathbf{f}$$

$$J_1\ddot{\phi} + \int_{-\infty}^{t} k_\theta(t-\tau)\frac{d}{d\tau}(\phi-\theta)\,d\tau + \sum_{i=1}^{n}\left[\mathbf{k}_i\cdot\left(\mathbf{u}-\mathbf{u}_i\right) + K_a\left(\phi+\phi_i\right) + K_l\left(\phi-\phi_i\right)\right] = g$$

$$M_2\ddot{\mathbf{v}} + \int_{-\infty}^{t} k_d(t-\tau)\frac{d}{d\tau}(\mathbf{v}-\mathbf{u})\,d\tau = \mathbf{0}$$

$$J_2\ddot{\theta} + \int_{-\infty}^{t} k_\theta(t-\tau)\frac{d}{d\tau}(\theta-\phi)\,d\tau = 0$$

(1)

where dot indicates derivative with respect to time $t$ and the auxiliary stiffness parameters

$$\mathbf{K}_i = E_s\left(\frac{w}{l}\right)\left[\left(\mathbf{d}_i\otimes\mathbf{d}_i\right) + \left(\frac{w}{l}\right)^2\left(\mathbf{t}_i\otimes\mathbf{t}_i\right)\right], \qquad \mathbf{k}_i = E_s\frac{a}{2}\left(\frac{w}{l}\right)^3\mathbf{t}_i,$$

$$K_a = E_s\left(\frac{w}{l}\right)^3\frac{a^2}{4}, \qquad K_l = E_s\left(\frac{w}{l}\right)^3\frac{l^2}{12}$$

(2)

while $k_d(t)$ and $k_\theta(t)$ are time-dependent relaxation functions accounting synthetically for the viscoelastic ring-resonator coupling. The in-plane displacement vector $\mathbf{u}_i$ and the rotation $\phi_i$ describe the motion of the $n$ rings connected to the reference ring. The unit vector $\mathbf{d}_i$ accounts for the orientation of the $i$-th connection ligament, and $\mathbf{t}_i$ is the unit vector normal to $\mathbf{d}_i$, according to a counter-clockwise system. The reference ring is excited by the generalized external forces **f** and $g$, while the resonator is assumed unloaded. It is worth noting that assuming time-independent relaxation functions allows the recovery of the dynamic equations governing the non-dissipative beam lattice metamaterial [53].

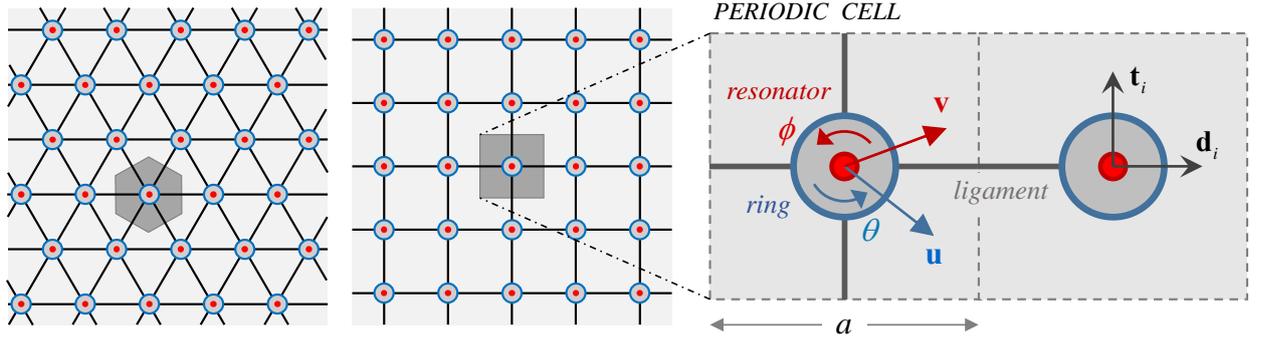

Figure 1. Beam lattice metamaterials and reference periodic cell.

The soft viscoelastic filler is characterized by the translational and rotational relaxation functions $k_d(t)$ and $k_\theta(t)$ that can be modeled by using the Prony series [42],[54],[55]. Considering only the first exponential term of the series, the relaxation functions read

$$k_d(t) = k_{de} + k_d \exp\left(-\frac{t}{t_r}\right) = k_{de}\left(1 + \beta_d \exp\left(-\frac{t}{t_r}\right)\right),$$
$$k_\theta(t) = k_{\theta e} + k_\theta \exp\left(-\frac{t}{t_r}\right) = k_{\theta e}\left(1 + \beta_\theta \exp\left(-\frac{t}{t_r}\right)\right),$$
(3)

where $t_r$ is the relaxation time, $k_{de}$, $k_d$, $k_{\theta e}$, $k_\theta$ are dimensional mechanical coefficients, and $\beta_d = k_d/k_{de}$, $\beta_\theta = k_\theta/k_{\theta e}$ are the associated nondimensional parameters, referred to as *viscosity ratios* in the following. The dependence of the nondimensional relaxation function $\tilde{k}_d = k_d/(E_s d)$ on the nondimensional time $\tilde{t} = t\sqrt{aE_S/M_1}$ is shown in Figure 2a tor a fixed relaxation time and different $\beta_d$-values. The increment of the exponential decay rate for increasing viscosity ratios $\beta_d$ can be appreciated. A similar qualitative scenario can be obtained for the time-dependence of the nondimensional relaxation function $\tilde{k}_\theta = k_\theta/(a^2 E_s d)$.

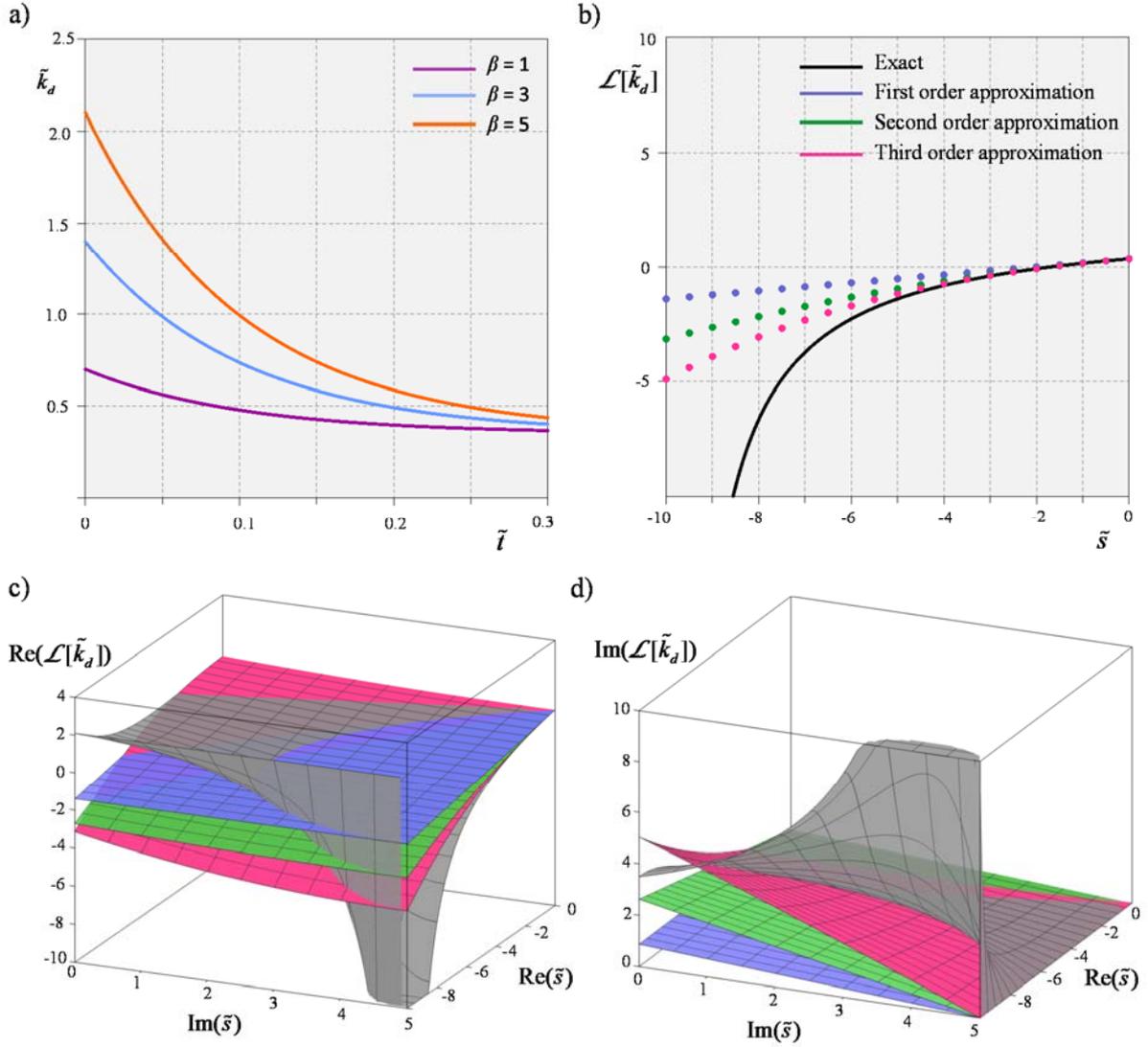

Figure 2. Viscoelastic kernel for $\tilde{t}_r = 1/10$, $\tilde{k}_{de} = k_{de}/(E_s d) = 35/100$: (a) Relaxation function versus time; (b) Exact and approximate Laplace transforms of the relaxation function versus the Laplace variable ($\beta_d = 5$), (c),(d) Real and imaginary parts of the Laplace transforms of the relaxation function versus the complex Laplace variable ($\beta_d = 5$).

## 3. In-plane Bloch waves

The propagation of elastic waves can be studied by applying the bilateral Laplace transform $\mathcal{L}[\bullet] = \int_{-\infty}^{\infty} (\bullet) \exp(-st) \, dt$ to the equations (1), where $s$ is the complex Laplace variable [56]. According to the Floquet-Bloch theory [57], the quasi-periodicity conditions can be imposed on the displacements and the rotation in the Laplace space

$$\mathcal{L}[\mathbf{u}_i] - \mathcal{L}[\mathbf{u}] = \hat{\mathbf{u}} \left( \exp(\imath \mathbf{k} \cdot \mathbf{x}_i) - 1 \right)$$
$$\mathcal{L}[\phi_i] \pm \mathcal{L}[\phi] = \hat{\phi} \left( \exp(\imath \mathbf{k} \cdot \mathbf{x}_i) \pm 1 \right), \quad (4)$$

where $\mathbf{x}_i$ is the vector pointing the $i$-th ring center, $\hat{\mathbf{u}}$, $\hat{\phi}$ are the displacement and rotation in the Bloch-Laplace space and $\mathbf{k} = q\mathbf{i}$ is the real-valued wavevector, with $q$ representing the wavenumber and $\mathbf{i}$ being the unit vector of the propagation direction. Therefore, the algebraic transformed equations of in-plane motion read

$$M_1 s^2 \hat{\mathbf{u}} + R_d(s)(\hat{\mathbf{u}} - \hat{\mathbf{v}}) + \sum_{i=1}^{n}\left[\mathbf{K}_i\left(1 - \exp(i\mathbf{k}\cdot\mathbf{x}_i)\right)\hat{\mathbf{u}} + \mathbf{k}_i\left(1 + \exp(i\mathbf{k}\cdot\mathbf{x}_i)\right)\hat{\phi}\right] = \hat{\mathbf{f}}$$

$$J_1 s^2 \hat{\phi} + R_\theta(s)\left(\hat{\phi} - \hat{\theta}\right) + \sum_{i=1}^{n}\left[\begin{matrix}\mathbf{k}_i \cdot \left(1 - \exp(i\mathbf{k}\cdot\mathbf{x}_i)\right)\hat{\mathbf{u}} + \\ + K_a\left(1 + \exp(i\mathbf{k}\cdot\mathbf{x}_i)\right)\hat{\phi} + K_l\left(1 - \exp(i\mathbf{k}\cdot\mathbf{x}_i)\right)\hat{\phi}\end{matrix}\right] = \hat{g} \qquad (5)$$

$$M_2 s^2 \hat{\mathbf{v}} + R_d(s)(\hat{\mathbf{v}} - \hat{\mathbf{u}}) = \mathbf{0}$$

$$J_2 s^2 \hat{\theta} + R_\theta(s)(\hat{\theta} - \hat{\phi}) = 0$$

where $\hat{\mathbf{f}}$ and $\hat{g}$ are the generalized forces in the Bloch-Laplace space. The two auxiliary $s$-dependent rational functions $R_d(s)$ and $R_\theta(s)$ are the bilateral Laplace transformation of the relaxation functions

$$R_d(s) = \mathcal{L}[k_d] = \frac{k_{de} + (k_{de} + k_d)st_r}{1 + st_r},$$

$$R_\theta(s) = \mathcal{L}[k_\theta] = \frac{k_{\theta e} + (k_{\theta e} + k_\theta)st_r}{1 + st_r}. \qquad (6)$$

If necessary for the sake of simplicity, the bilateral Laplace transformations (6) can be approximated with their $h$-order Taylor polynomials, centered at $s = 0$, yielding

$$R_d(s) \simeq k_{de} - k_d \sum_{j=1}^{h}(-1)^j t_r^j s^j$$

$$R_\theta(s) \simeq k_{\theta e} - k_\theta \sum_{j=1}^{h}(-1)^j t_r^j s^j \qquad (7)$$

It is worth noting that for $h = 1$ the Taylor polynomials introduce in the equations (5) some linear terms in the Laplace variable, recovering the classical viscous damping originated by velocity-proportional dissipation. Specifically, the first, second and third order polynomial approximations of the Laplace transformations (6) are reported in Figure 2b considering $\mathcal{L}[\tilde{k}_d]$ as an analytical function of the nondimensional variable $\tilde{s} = s\sqrt{M_1/(aE_s)}$, fixed a certain nondimensional relaxation time $\tilde{t}_r$. The comparison with the exact function allows to appreciate the approximation accuracy in the closeness of the starting point $\tilde{s} = 0$ up to the polar singularity at $\tilde{s} = 1/\tilde{t}_r$. Furthermore, the real part $\text{Re}\left(\mathcal{L}[\tilde{k}_d]\right)$ and the imaginary part $\text{Im}\left(\mathcal{L}[\tilde{k}_d]\right)$ are properly compared with the respective first, second and third order approximations as a function of the complex $\tilde{s}$-variable in Figure 2c,d.

The equivalent matrix form of the linear algebraic equations (5) is $\mathbf{C}(\mathbf{k},s)\hat{\mathbf{U}} = \hat{\mathbf{F}}$, where $\hat{\mathbf{U}} = \begin{pmatrix} \hat{\mathbf{u}} & \hat{\phi} & \hat{\mathbf{v}} & \hat{\theta} \end{pmatrix}^T$, $\hat{\mathbf{F}} = \begin{pmatrix} \hat{\mathbf{f}} & \hat{g} & \mathbf{0} & 0 \end{pmatrix}^T$ and the matrix $\mathbf{C}(\mathbf{k},s)$ is a 6-by-6 Hermitian matrix can be written

$$\mathbf{C}(\mathbf{k},s) = \begin{bmatrix} \mathbf{A} + s^2 M_1 \mathbf{I} & \mathbf{a}^- & -R_d(s)\mathbf{I} & \mathbf{0} \\ \mathbf{a}^+ & b + s^2 J_1 & \mathbf{0} & -R_\theta(s) \\ -R_d(s)\mathbf{I} & \mathbf{0} & R_d(s)\mathbf{I} + s^2 M_2 \mathbf{I} & \mathbf{0} \\ \mathbf{0} & -R_\theta(s) & \mathbf{0} & R_\theta(s) + s^2 J_2 \end{bmatrix} \quad (8)$$

where $\mathbf{I}$ is the 2-by-2 identity matrix and the submatrix $\mathbf{A}$, $\mathbf{a}^-$, $\mathbf{a}^+$ and $b$ are

$$\mathbf{A} = R_d(s)\mathbf{I} + \sum_{i=1}^{n} \left[ (1 - \exp(\imath \mathbf{k} \cdot \mathbf{x}_i)) \mathbf{K}_i \right] = R_d(s)\mathbf{I} + \sum_{i=1}^{n} \left[ (1 - \cos(\mathbf{k} \cdot \mathbf{x}_i)) \mathbf{K}_i \right]$$

$$\mathbf{a}^- = \sum_{i=1}^{n} \left[ (1 + \exp(\imath \mathbf{k} \cdot \mathbf{x}_i)) \mathbf{k}_i \right] = -\imath \sum_{i=1}^{n} \left[ \sin(\mathbf{k} \cdot \mathbf{x}_i) \mathbf{k}_i \right]$$

$$\mathbf{a}^+ = \sum_{i=1}^{n} \left[ (1 - \exp(\imath \mathbf{k} \cdot \mathbf{x}_i)) \mathbf{k}_i^T \right] = \imath \sum_{i=1}^{n} \left[ \sin(\mathbf{k} \cdot \mathbf{x}_i) \mathbf{k}_i^T \right] \quad (9)$$

$$b = R_\theta(s) + \sum_{i=1}^{n} \left[ K_a (1 + \exp(\imath \mathbf{k} \cdot \mathbf{x}_i)) + K_l (1 - \exp(\imath \mathbf{k} \cdot \mathbf{x}_i)) \right] =$$

$$= R_\theta(s) + \sum_{i=1}^{n} \left[ K_a (1 + \cos(\mathbf{k} \cdot \mathbf{x}_i)) + K_l (1 - \cos(\mathbf{k} \cdot \mathbf{x}_i)) \right]$$

where the equivalent exponential and trigonometric forms are reported.

## 4. Free wave propagation

For the free wave propagation, the complex-valued dispersion relations $s(\mathbf{k})$ can be determined by solving the eigenvalue problem associated to the homogeneous equations of motion $\mathbf{C}(\mathbf{k},s)\hat{\mathbf{U}} = \mathbf{0}$, obtained by zeroing the generalized external forces ($\hat{\mathbf{f}} = \mathbf{0}$, $\hat{g} = 0$). As preliminary remark, it must be highlighted that different eigenvalue problems are associated to the exact relaxation functions (6) and to each order of their approximations in $s$-power series (7). Specifically, the exact and approximate relaxation functions correspond to rational (non-polynomial) and polynomial eigenvalue problems in the $s$-unknown, respectively. An algebraic procedure to simplify the mathematical treatment of the eigenvalue problems (by conveniently reducing both rational and polynomial problems to higher dimension linear problems) is reported in Appendix A.

The dispersion spectrum is composed by all the complex-valued relations $s(\mathbf{k})$ solving the eigenvalue problems for assigned real $\mathbf{k}$-values. Precisely, the real and imaginary parts of each solution define a pair of dispersion surfaces defined in the bi-dimensional $\mathbf{k}$-domain. Alternately, the dispersion spectrum can be illustrated by representing the three-dimensional dispersion curves (or spectral branches) defined by the real and imaginary parts $\mathrm{Re}(s)$ and $\mathrm{Im}(s)$ of the complex-valued relations $s(q)$ along a particular propagation direction.

Focusing on the cellular metamaterial characterized by the quadrilateral beam lattice, the eigenvalue problem is stated for the particular periodic cell with square shape ($a = l + 2R$). For this metamaterial the submatrices (9) assume the particular form

$$\mathbf{A} = \begin{bmatrix} R_d(s) + \dfrac{2E_s w C(k_1)}{l} + \dfrac{2E_s w^3 C(k_2)}{l^3} & 0 \\ 0 & R_d(s) + \dfrac{2E_s w^3 C(k_1)}{l^3} + \dfrac{2E_s w C(k_2)}{l} \end{bmatrix},$$

$$\mathbf{a}^- = \begin{bmatrix} -\dfrac{\iota E_s w^3 S(k_2)}{l^2} & \dfrac{\iota E_s w^3 S(k_1)}{l^2} \end{bmatrix}^T, \tag{10}$$

$$\mathbf{a}^+ = \begin{bmatrix} \dfrac{\iota E_s w^3 S(k_2)}{l^2} & -\dfrac{\iota E_s w^3 S(k_1)}{l^2} \end{bmatrix},$$

$$b = R_\theta(s) + \dfrac{E_s w^3}{3l}\left[2 + C(k_1) + C(k_2)\right],$$

where the auxiliary trigonometric functions $C(k_i) = 1 - \cos(k_i a)$ and $S(k_i) = \sin(k_i a)$, with $i=1,2$. The analysis of the dispersion spectrum can be focused on the first Brillouin zone $\mathcal{B}$ of the bi-dimensional $\mathbf{k}$-domain [57], which is defined $\mathcal{B} = \{\mathbf{k}: k_i a = \tilde{k}_i \in [-\pi, \pi]\}$ in nondimensional form for the quadrilateral metamaterial. Within this square zone, the complex-valued spectrum can be synthetically illustrated by determining the dispersion curves along particular propagation directions. Specifically, the real and imaginary parts of all the dispersion curves can be described over the closed boundary $\partial \mathcal{B}$ of the triangular subdomain $\mathcal{B}_1 \subset \mathcal{B}$, limited by the vertices $B_1, B_2, B_3$ (pointed by the nondimensional wavevectors $\tilde{\mathbf{k}}_1 = (0\ 0)^T, \tilde{\mathbf{k}}_2 = (0\ \pi)^T, \tilde{\mathbf{k}}_3 = (\pi\ \pi)^T$ respectively). Accordingly, the boundary $\partial \mathcal{B}$ is spanned by the nondimensional curvilinear abscissa $\xi$, varying in the range $[0, 2\pi + \sqrt{2}\pi]$.

### 4.1. Complex-valued dispersion spectra

Focusing first on the lowest (first) order approximation of the relaxation functions, the dispersion curves related to the particular metamaterial $\mathcal{M}$ are portrayed in Figure 3. The selected mechanical parameters are $\tilde{R} = R/a = 1/5$, $\tilde{w} = w/a = 3/50$, $\tilde{r} = r/a = 1/10$, $\tilde{J}_1 = J_1/(M_1 a^2) = \tilde{R} - \tilde{R}\tilde{w} + \tfrac{1}{2}\tilde{w}^2$, $\tilde{M}_2 = M_2/M_1 = \tilde{\rho}\tilde{r}^2/(2\tilde{R}\tilde{w} - \tilde{w}^2)$, $\tilde{J}_2 = J_2/(M_1 a^2) = \tfrac{1}{2}\tilde{\rho}\tilde{M}_2\tilde{r}^2$

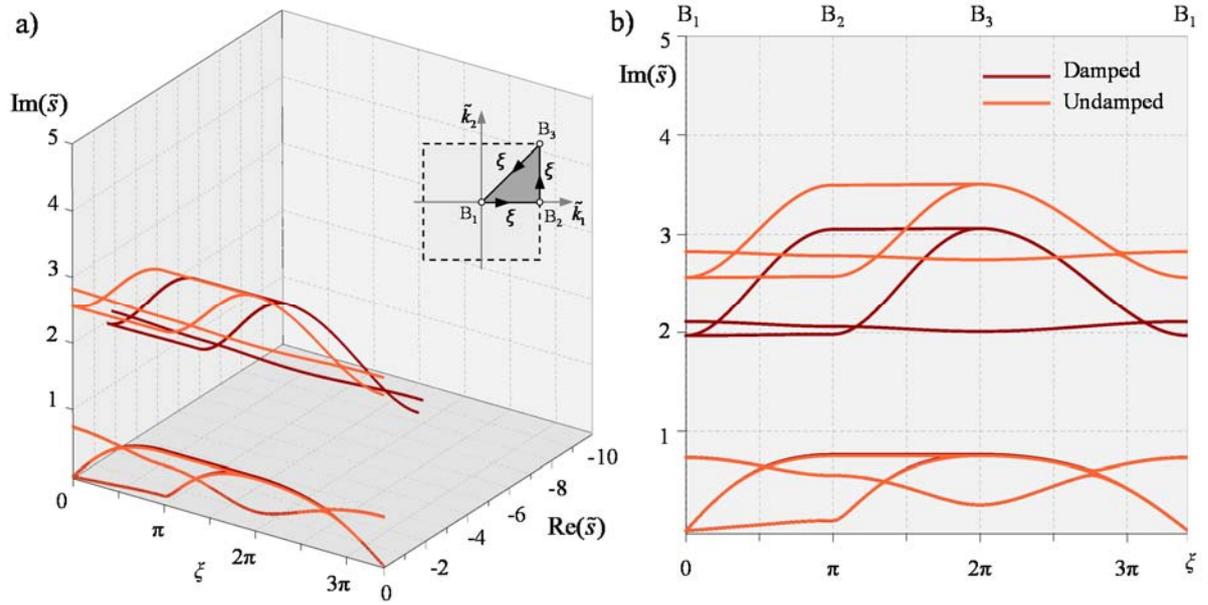

Figure 3. Complex-valued dispersion spectrum of the metamaterial $\mathcal{M}$ corresponding to the first-order approximation of the relaxation functions: (a) wave frequency and wave damping, (b) wave frequency.

and resonator-to-ring mass density ratio $\tilde{\rho}=10$. Dispersion curves related to visco-elastically damped resonators ($\tilde{k}_{de}=35/100$, $\tilde{k}_{\theta e}=16/2500$, $\beta=\beta_d=\beta_\theta=5$, $\tilde{t}_r=1/10$) and undamped resonators ($\beta=\beta_d=\beta_\theta=0$ or $\tilde{t}_r\to\infty$) are illustrated.

The real and imaginary parts of the complex frequency $\tilde{s}$, which can be referred to as wave *damping* and wave *frequency* in the following, are related to the propagation in space and the attenuation in time of the mono-harmonic wave traveling through the metamaterial. Negative real parts of the complex frequency correspond to exponentially decaying amplitudes of the propagating wave. As expected, the undamped metamaterial shows a dispersion diagram composed by six purely imaginary dispersion curves, corresponding to waves propagating without attenuation (Figure 3a). A full large-amplitude band gap separates the three low-frequency dispersion curves from the three high-frequency dispersion curves (Figure 3b). It can be verified that the low-frequency curves are associated to waveforms systematically localized in the generalized ring displacements (*ring polarization*), with quasi-static contribution of the generalized resonator displacements. Differently, the high-frequency curves are associated to waveforms mainly localized in the generalized resonator displacements (*resonator polarization*). Looking at the damped metamaterial, the dispersion spectrum possesses six complex-valued curves (Figure 3a). The three curves in the low frequency range can be conventionally referred to as spectral branches of *quasi-propagation*, since they are dominated by the imaginary part of the complex frequency, with minimal participation of the real part. It can be verified that the slight dynamic interaction between the ring and the resonator, caused by the ring polarization, reduces the attenuation offered by the viscoelastic coupling. Consequently, no appreciable differences can be recognized in the low-frequency spectra of the damped and undamped metamaterials. The remaining three curves in the high frequency range can be conventionally referred to as spectral branches of *strong-attenuation*, since they are significantly contributed by the real part of the complex frequency. It can be verified that the strong ring-resonator interaction caused by the resonator polarization increases the attenuation offered by the viscoelastic coupling. As important remark, the strong frequency reduction caused by the viscoelastic coupling in the damped metamaterial causes a marked decrement in the band gap amplitude (Figure 3b).

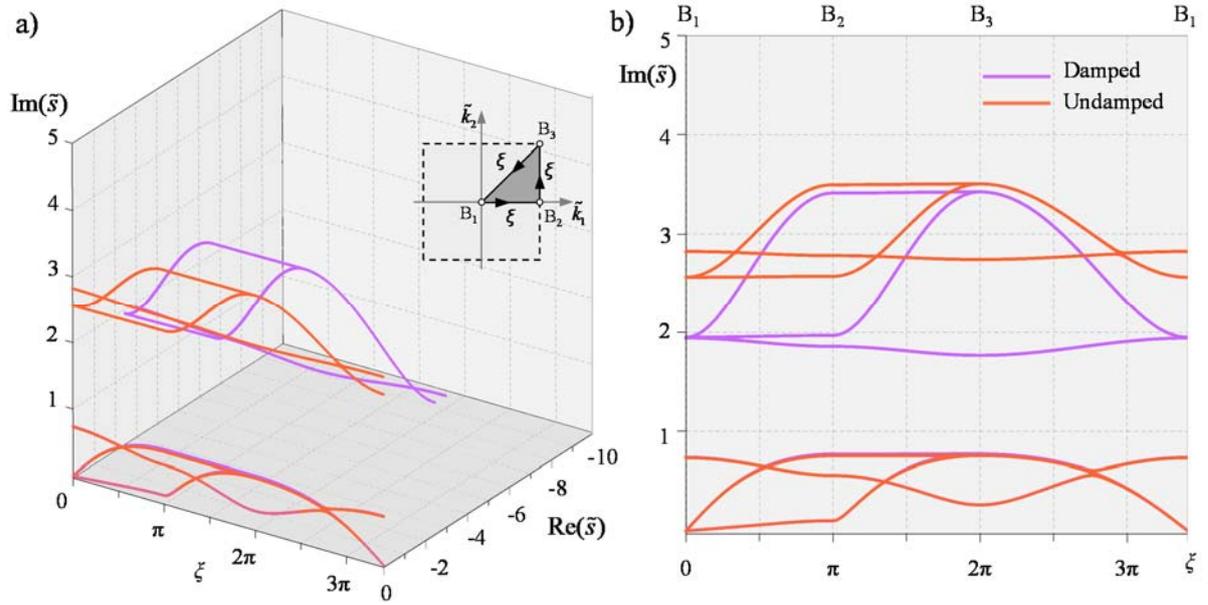

Figure 4. Complex-valued dispersion spectrum of the metamaterial $\mathcal{M}$ corresponding to the second-order approximation of the relaxation functions: (a) wave frequency and wave damping, (b) wave frequency.

Considering a second-order approximation of the relaxation functions, the approximating series introduce $\tilde{s}^2$-proportional contributions to the polynomial eigenvalue problem, which can modify the second $\tilde{s}$-power terms deriving from the inertia forces. The corresponding dispersion curves are portrayed in Figure 4. As first remark, the real part of all the dispersion curves tends to decrease, causing a small increment of the damping in the propagating waves (Figure 4a). As major remark, the essential difference with respect to the first-order approximation is related to the high-frequency dispersion curves, which concur to determine a pass band with significantly larger amplitude (Figure 4b). The amplitude enlargement also slightly reduces the band gap between the high frequency and the low frequency branches of the spectrum. As complementary remark, it can be noted that the limit of long wavelengths ($\xi = 0$) is a triple frequency point for the high-frequency dispersion curves.

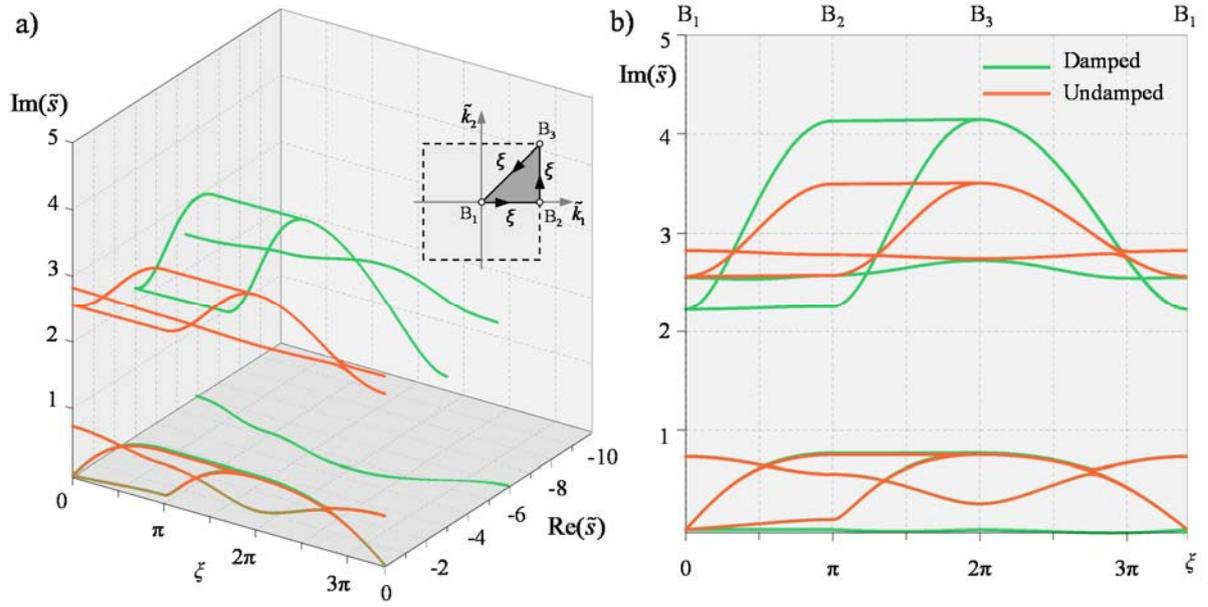

Figure 5. Complex-valued dispersion spectrum of the metamaterial $\mathcal{M}$ corresponding to the third-order approximation of the relaxation functions: (a) wave frequency and wave damping, (b) wave frequency.

Considering a third-order approximation of the relaxation functions, the approximating series introduce new $\tilde{s}^3$-proportional contributions to the polynomial eigenvalue problem. The corresponding dispersion curves are portrayed in Figure 5. As first remark, the real part of all the dispersion curves shows a further decrement (damping increment) with respect to the second order approximation (Figure 5a). As major remark, the order augment of the polynomial eigenvalue problem determines the birth of a new real-valued dispersion curve of the spectrum. Consequently, although the number of non-zero wave frequencies remains unchanged, the number of branches in the complex-valued dispersion spectrum exceeds the total number of degrees-of-freedom in the periodic cell. It is worth remarking that the new real-valued dispersion curve corresponds to waves non-propagating in space but highly damped in time. As minor remark, the pass band associated to the high-frequency dispersion curves is shifted to a higher frequency range and also increases in amplitude with respect to the second order approximation (Figure 5b).

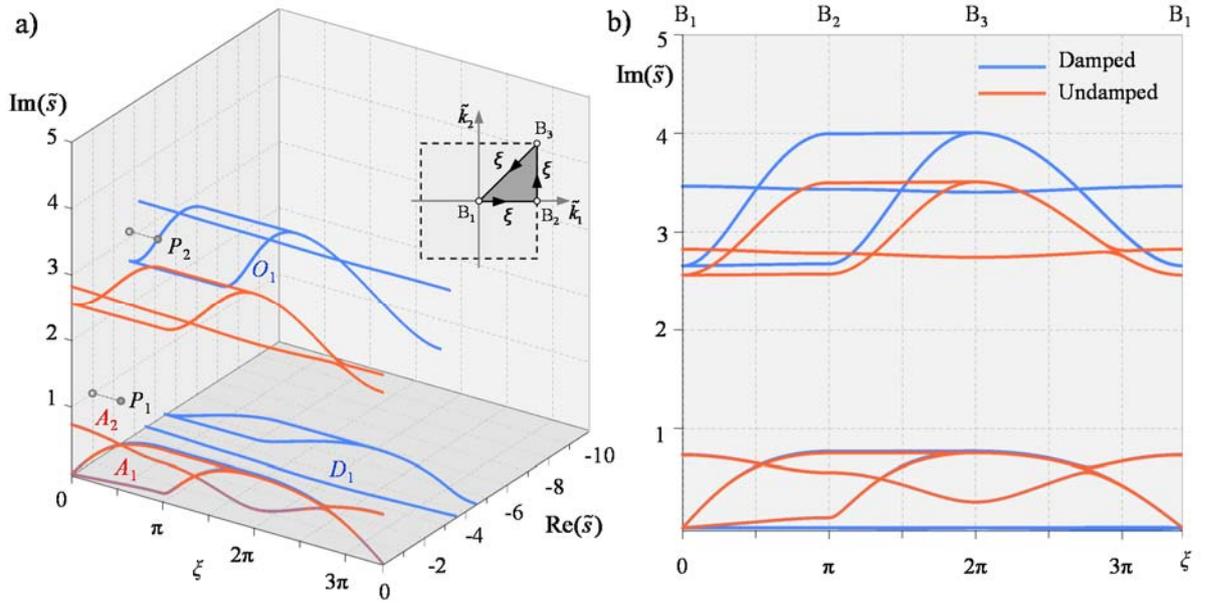

Figure 6. Complex-valued dispersion spectrum of the metamaterial $\mathcal{M}$ corresponding to the exact relaxation functions: (a) wave frequency and wave damping, (b) wave frequency.

Finally, considering the exact relaxation functions, the wave dispersion properties are governed by a rational eigenvalue problem. The corresponding dispersion curves are shown in Figure 6. As for the third-order approximation, the number of spectrum branches exceeds the total number of degrees-of-freedom in the periodic cell. Specifically, the spectrum shows three real-valued dispersion curve corresponding to non-propagating damped waves (Figure 6a). In the comparison with the undamped metamaterial, it can be clearly recognized that the exact treatment of the viscoelastic coupling determines a slight amplification of the band gap separating the low-frequency and the high-frequency dispersion curve (Figure 6b). The relevance of this key observation is principally related to the qualitative and quantitative comparison with the approximate (first-order) treatment of the viscoelastic coupling. Indeed, from the quantitative viewpoint, the classical first-order approximation is found to strongly underestimate the band gap amplitude. Furthermore, from the qualitative viewpoint, it also returns a band gap reduction with respect to the undamped metamaterial. As complementary remark, the amplitude of the high-frequency pass band is also amplified. For the sake of completeness, a wide parametric analyses has been performed to assess the effect of different viscosity ratios on the exact dispersion spectrum. To exemplify the results, the dispersion spectrum related to a smaller viscosity ratio ( $\beta = 3$ ) is shown in Figure 7. As major remark, the lower viscosity causes a systematic augment of the real part (damping reduction) for all the dispersion curves associated to propagating waves. On the contrary, the lower viscosity systematically shifts the dispersion curves associated to non-propagating waves to a lower real value range (Figure 7a). Smaller viscosity ratios also tends to reduce the amplitudes of the band gap (Figure 7b).

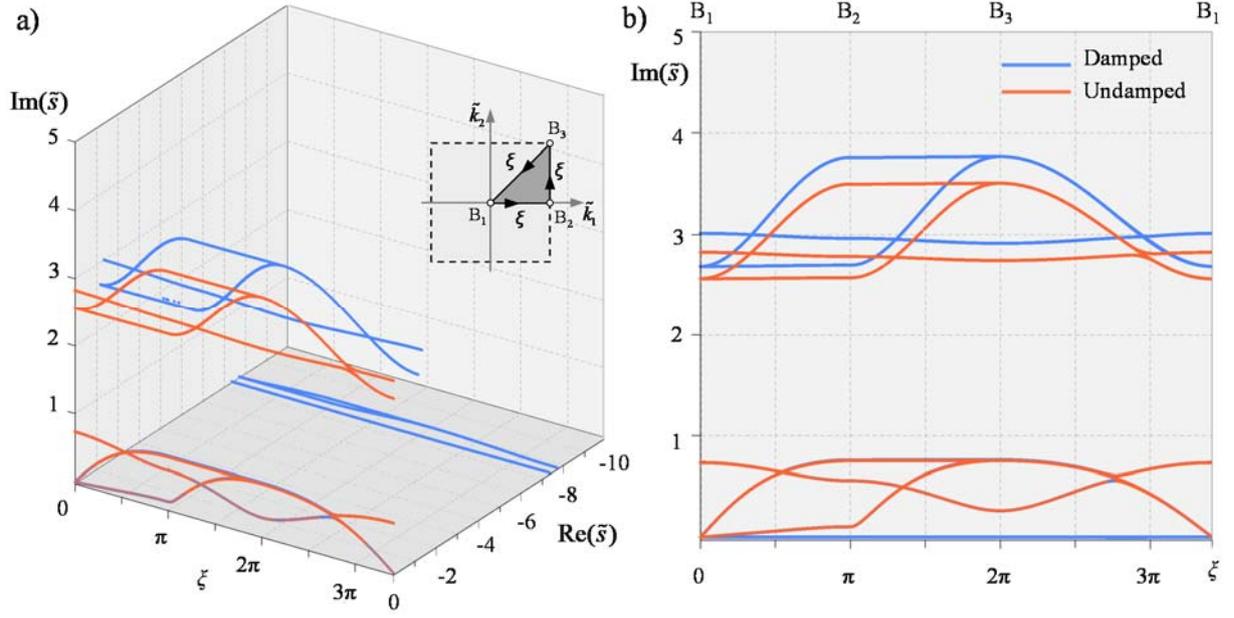

Figure 7. Complex-valued dispersion spectrum of the metamaterial $\mathcal{M}$ (with low viscosity ratio $\beta = 3$) corresponding to the exact relaxation functions: (a) wave frequency and wave damping, (b) wave frequency.

## 5. Forced wave propagation

For time-harmonic forced waves, the generalized external forces applied at the generic ring pointed by the position vector $\mathbf{x}_n$ can be expressed as $\mathbf{f}_n = \mathbf{p}\, H(t)\exp(St)\exp(\iota \mathbf{K}\cdot \mathbf{x}_n)$ and $g_n = m\, H(t)\exp(St)\exp(\iota \mathbf{K}\cdot \mathbf{x}_n)$, where $H(t)$ is the unit step function, $\mathbf{p}$ and $m$ are time-independent force amplitudes, $S$ is the complex-valued forcing frequency and $\mathbf{K}$ is the real-valued forcing wavevector. By virtue of the lattice periodicity, the application point at the reference ring (at position $\mathbf{x}_n = \mathbf{0}$) can be studied without loss of generality. Therefore, the right-hand term of the equations (1) governing the forced dynamics of the reference ring read

$$\mathbf{f}(S,t) = \mathbf{p}\, H(t)\exp(St)$$
$$g(S,t) = m\, H(t)\exp(St) \tag{11}$$

where the real part of the complex-valued frequency $S$ is assumed not null and negative in the general case, in order to account for generic, exponentially decaying external loads.

Consequently, since the linear algebraic equations (5) have been written in the matrix form $\mathbf{C}(\mathbf{k},s)\hat{\mathbf{U}} = \hat{\mathbf{F}}(S,s)$, the right-hand term $\hat{\mathbf{F}}(S,s) = \begin{pmatrix} \hat{\mathbf{f}}(S,s) & \hat{g}(S,s) & 0 & 0 \end{pmatrix}^T$ is obtained by applying the bilateral Laplace transform to equation (11), yielding

$$\hat{\mathbf{f}}(S,s) = \frac{\mathbf{p}}{s-S}$$
$$\hat{g}(S,s) = \frac{m}{s-S} \qquad (12)$$

where $s = S$ can be easily recognized as a simple pole of the transformed external forces.

Assigned a generic forcing frequency $S$, the stationary lattice response is described by the transformed displacement vector in the Bloch-Laplace space

$$\hat{\mathbf{U}}(S,\mathbf{k},s) = \mathbf{D}(\mathbf{k},s)\hat{\mathbf{F}}(S,s) \qquad (13)$$

where the six-by-six matrix $\mathbf{D}(\mathbf{k},s) = \mathbf{C}(\mathbf{k},s)^{-1}$ is also known as dynamic *compliance* matrix. According to this formulation, the roots of the characteristic equations $\det \mathbf{C}(\mathbf{k},s) = 0$, that define the Floquet-Bloch spectrum of the metamaterial, are expected to determine poles in the components of the compliance matrix $\mathbf{D}(\mathbf{k},s)$. From the mechanical viewpoint, the forced response amplitude is essentially determined by the relative position between these spectral poles and the poles of the transformed external forces.

Therefore, accordingly with [58], the transformed displacement vector $\hat{\mathbf{U}}(S,\mathbf{k},s)$ of the reference cell can be integrated in the entire first (dimensional) Brillouin zone $\mathcal{D} = \{\mathbf{k} : k_i \in [-\pi/a, \pi/a]\}$ by means of the Fourier integral

$$\hat{\mathbf{U}}(S,s) = \frac{1}{4\pi^2}\int_{\mathcal{D}} \hat{\mathbf{U}}(S,\mathbf{k},s)\,d\mathbf{k} = \frac{1}{4\pi^2}\int_{\mathcal{D}} \mathbf{D}(\mathbf{k},s)\hat{\mathbf{F}}(S,s)\,d\mathbf{k} \qquad (14)$$

where $\hat{\mathbf{U}}(S,s)$ expresses the transformed displacement vector in the Laplace space, corresponding also to the inverse space Fourier transform of $\hat{\mathbf{U}}(S,\mathbf{k},s)$.

The stationary lattice response is finally expressed in the time-dependent complex-valued displacement vector $\mathbf{U}(S,t)$ of the reference cell (centered at position $\mathbf{x}_n = \mathbf{0}$), by applying the inverse bilateral Laplace transform

$$\mathbf{U}(S,t) = \mathcal{L}^{-1}\left[\hat{\mathbf{U}}(S,s)\right] = \frac{1}{2\pi \iota}\int_{r-\iota\infty}^{r+\iota\infty} \hat{\mathbf{U}}(S,s)\exp(st)\,ds, \qquad r \in \mathbb{R} \qquad (15)$$

which is evaluated as $\mathbf{U}(S,t) = \sum R\left(\hat{\mathbf{U}}(S,s)\exp(st)\right)$, where $R$ stands for the residual of $\hat{\mathbf{U}}(S,s)\exp(st)$ and the sum is extended to all the poles of $\hat{\mathbf{U}}(S,s)$. Finally the time-dependent displacement vector $\mathbf{U}_n(\mathbf{K},S,t)$ of the generic cell centered at the point $\mathbf{x}_n$ can be determined by applying the relation $\mathbf{U}_n(\mathbf{K},S,t) = \mathbf{U}(S,t)\exp(\imath\mathbf{K}\cdot\mathbf{x}_n)$.

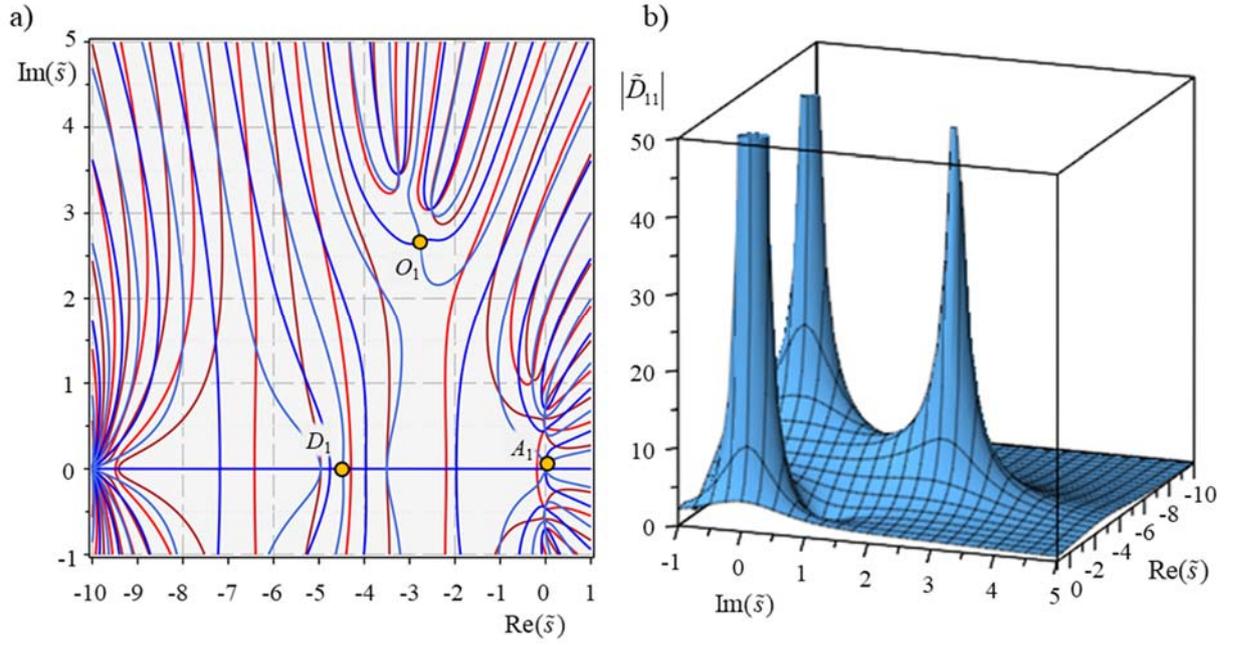

Figure 8. Compliance matrix of the particular metamaterial $\mathcal{M}$ : (a) loci of the null real and null imaginary parts of the $\tilde{D}_{11}$-numerator (curves in red-scale) and $\tilde{D}_{11}$-denominator (curves in blue-scale), (b) magnitude of $\tilde{D}_{11}$.

## 5.1 Compliance matrix

In order to determine and discuss the forced response of the particular metamaterial $\mathcal{M}$, the dynamic compliance matrix can be analyzed first. The nondimensional complex-valued component $\tilde{D}_{11}(\tilde{\mathbf{k}}_0,\tilde{s}) = D_{11}(\tilde{\mathbf{k}}_0,\tilde{s})E_s$ is considered for a fixed nondimensional wavevector $\tilde{\mathbf{k}}_0$. Selecting the particular wavevector $\tilde{\mathbf{k}}_0 = (1\ 0)^T$, the continuous loci of the null real and null imaginary parts of the $\tilde{D}_{11}$-denominator (coincident with the $\tilde{\mathbf{C}}(\tilde{\mathbf{k}}_0,\tilde{s})$-determinant) are portrayed in Figure 8a (curves in blue-scale) in the plane of the real and imaginary part of the complex frequency $\tilde{s}$. In the general case, the crossing points of the loci do not identify points of polar singularities or poles for the dynamic compliance, since they can coincide with intersections between the continuous loci of the null real and null imaginary parts of the $\tilde{D}_{11}$-numerator (curves in red-scale). On the contrary, the poles for the dynamic compliance are identified by the few crossing points in which the denominator vanishes for non-null values of the numerator. These peculiar points are characteristic properties of the metamaterial and can be referred to as *complex resonance points*. For the particular $\tilde{D}_{11}$-component under investigation, three distinct resonance points can be recognized (marked by yellow circles). The magnitude of the complex-valued component $\tilde{D}_{11}(\tilde{\mathbf{k}}_0,\tilde{s})$ is found to rapidly but continuously grow up in the closeness of these points, approaching infinite values (Figure 8b). It can be remarked that the resonance points actually coincide with some of the $\tilde{s}$-values already identified by the dispersion function $\tilde{s}(\tilde{\mathbf{k}}_0)$ in the free vibration analysis. In particular, the resonance points can be associated to the first acoustic branch (point $A_1$), to one of the purely damping branches (point $D_1$), and to one of the optical branches (point $O_1$) of the complex-valued spectrum shown in Figure 6a. Similarly, resonance points associable to the same or some other branches of the metamaterial spectrum can be found for each component of the compliance matrix.

## 5.2 Forced response

Focusing on the transformed displacements of the metamaterial $\mathcal{M}$ under the effect of the external complex-valued exponentially decaying force $\hat{\mathbf{F}}(S_0,s)$ with fixed forcing frequency $S_0$, the wavevector $\tilde{\mathbf{k}}$-dependence of the complex-valued displacement component $\hat{U}_1$ is firstly analyzed. According to equation (13), the displacement component is expressed by the linear combination $\hat{U}_1 = D_{11}\hat{F}_1 + D_{12}\hat{F}_2 + D_{13}\hat{F}_3$. In Figure 9, the magnitude of the nondimensional component $\tilde{\hat{U}}_1 = \hat{u}_1\sqrt{E_s/(M_1 a)}$ is reported over the first square Brillouin zone $\mathcal{B}$ for the particular nondimensional forcing frequency $\tilde{S}_0 = 3\iota$ and for unitary values of the nondimensional force amplitudes. Choosing different values of the complex frequency $\tilde{s}$, strongly dissimilar values of the $\tilde{\hat{U}}_1$-magnitude are obtained, depending on whether the frequency $\tilde{s}$ falls within the stop band ($\tilde{s}_1 = -1+\iota$, see Figure 9a) or close to a dispersion curve within a pass band ($\tilde{s}_2 = -2.51+3.05\iota$, see Figure 9b). As first remark, the magnitude of the linear combination $\tilde{\hat{U}}_1$ is geometrically non-symmetric in the square Brillouin zone $\mathcal{B}$, in the general case. However, the cubic symmetry of the metamaterial microstructure can be verified to systematically determine doubly symmetric magnitudes of each term contributing to the linear combination. As major remark, when the frequency $\tilde{s}$ cannot resonates because it falls within the stop band ($\tilde{s}_1$ associated to point $P_1$ in Figure 6a), the metamaterial response does not show any evident peak in the entire Brillouin zone $\mathcal{B}$, independently of the wavevector $\tilde{\mathbf{k}}$ (Figure 9a). On the contrary, when the frequency falls within the pass band ($\tilde{s}_2$ associated to point $P_2$ in Figure 6a), the metamaterial response can reach infinite-valued amplitudes, depending on whether the $(\tilde{s},\tilde{\mathbf{k}})$-combination lies on one of the dispersion curves (Figure 9b). Indeed, infinite-valued amplitudes are obtained for the frequency $\tilde{s}_2$ at resonance, that is in the closeness of the wavenumber $\tilde{k}_1 = 1$ (Figure 9c), which exactly corresponds to a dispersion curve of the spectrum (point $P_2$ in Figure 6a) and also to one of the singularities (point $O_1$) of the compliance matrix coefficient $\tilde{D}_{11}$ (portrayed in Figure 8 for $\tilde{\mathbf{k}}_0 = (1\ 0)^T$ or $\xi = 1$ in Figure 6a). On the contrary, if a slight variation is introduced to shift the frequency $\tilde{s}$ from the dispersion curve (quasi-resonance), the metamaterial response reaches high but finite-valued amplitudes (see for instance the $\hat{\tilde{U}}_1$-magnitude shown in Figure 9d for $\tilde{s}_3 = 1.005\ \tilde{s}_2$).

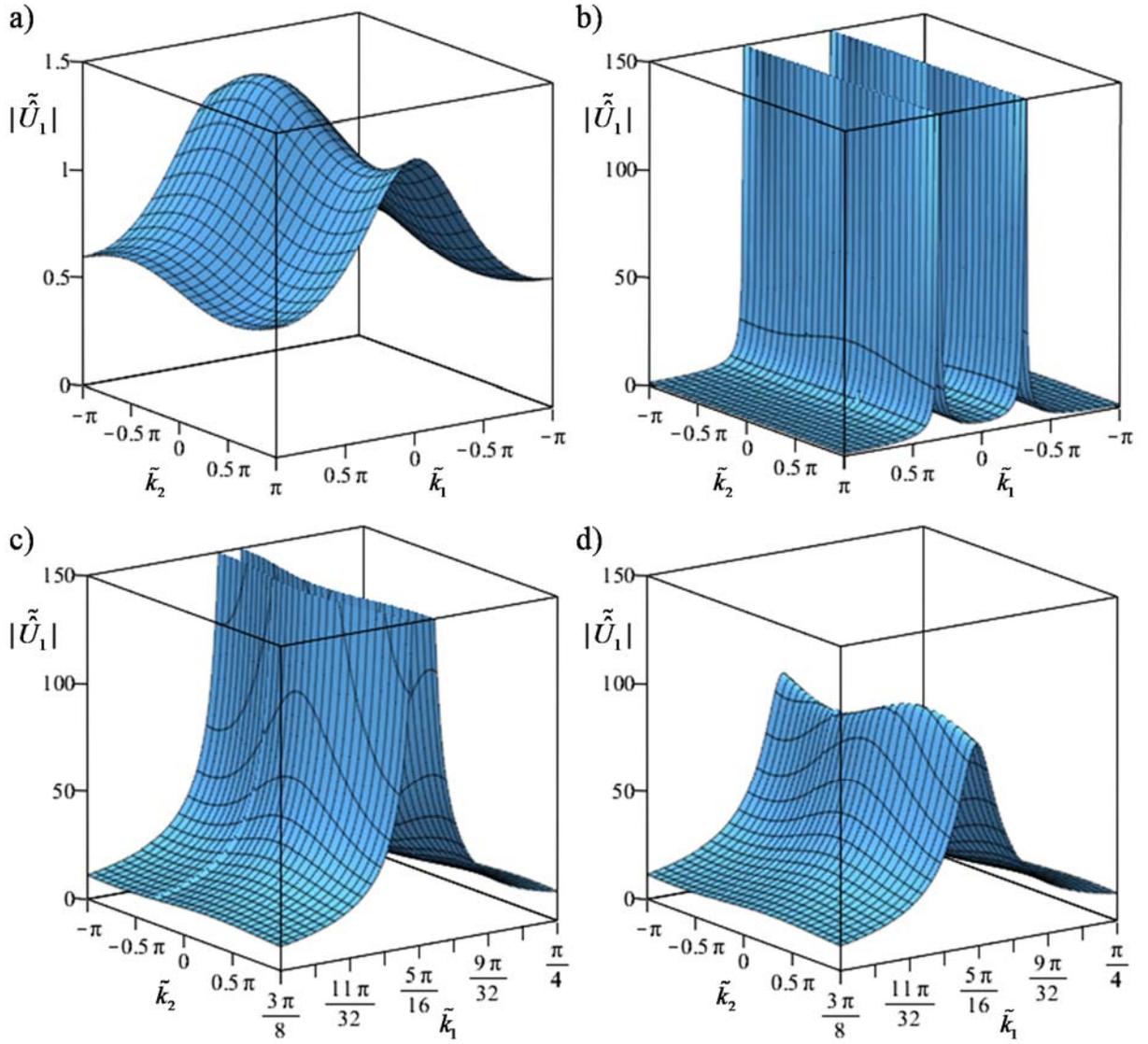

Figure 9. Magnitude of the displacement component $\hat{\tilde{U}}_1$ in the forced response of the particular metamaterial $\mathcal{M}$: (a) not-resonant complex frequency $\tilde{s}_1 = -1 + \iota$ falling within the stopband, (b),(c) complex frequency $\tilde{s}_2 = -2.51 + 3.05\iota$ falling within the passband at resonance, (d) complex frequency $\tilde{s}_3 = 1.005\ \tilde{s}_2$ falling within the passband at quasi-resonance.

Therefore, attention is focused on the frequency $\tilde{s}$-dependence of the complex-valued displacement component $\hat{U}_1$ of the metamaterial $\mathcal{M}$, which is again analyzed under the effect of the external complex-valued exponentially decaying force $\hat{\mathbf{F}}(S_0, s)$ with fixed forcing frequency $S_0$. In Figure 10, the nondimensional component $\hat{\tilde{U}}_1$ is reported versus the real and imaginary parts of the frequency $\tilde{s}$ for the wavevector $\tilde{\mathbf{k}}_0 = \begin{pmatrix} 1 & 0 \end{pmatrix}^T$, if the external forces are characterized by the forcing frequency $\tilde{S}_0 = 3\iota$ and unitary amplitudes. In this case the forcing frequency does not resonate with any of the metamaterial frequencies. The continuous loci of the null real and null imaginary parts of the $\hat{U}_1$-denominator are portrayed in Figure 10a (curves in blue-scale). All the crossing points of the loci that do not identify poles, because they also coincide with intersections between the continuous loci of the null real and null imaginary parts of the $\hat{U}_1$-numerator (curves in red-scale), are not significant. The remaining crossing points identify poles of the forced response. Among them, three poles (points $D_2, O_2, A_2$ marked by yellow dots) are characteristic properties of the metamaterial, while a fourth pole (point $G_1$ marked by green dot) is associated to the forcing frequency. It is worth noting that the $\hat{U}_1$-poles do not correspond exactly to the $\tilde{D}_{11}$-poles in Figure 8a, due to the second and third contributions to the linear combination $\hat{U}_1 = D_{11}\hat{F}_1 + D_{12}\hat{F}_2 + D_{13}\hat{F}_3$. The magnitude of the complex-valued component $\hat{U}_1$ is found to rapidly but continuously grow up in the closeness of all the four poles (and their complex conjugates), approaching infinite values (Figure 10b).

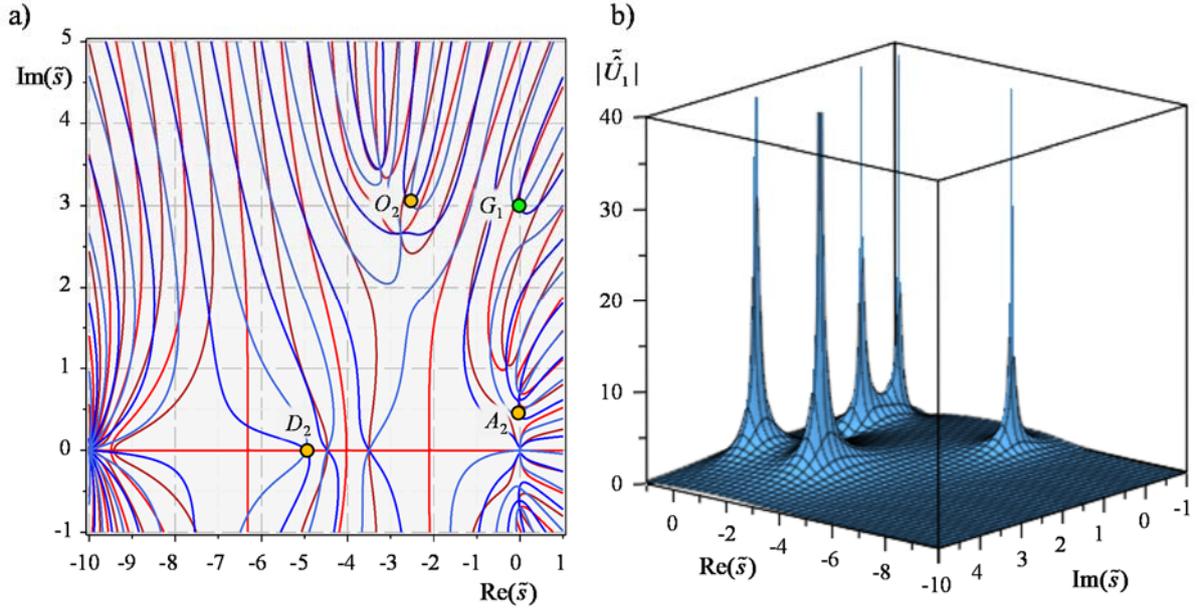

Figure 10. Response of the particular metamaterial $\mathcal{M}$ to non-resonant force ($\tilde{S}_1 = 3\iota$): (a) loci of the null real and null imaginary parts of the displacement component $\hat{\tilde{U}}_1$-numerator (curves in red-scale) and $\hat{\tilde{U}}_1$-denominator (curves in blue-scale), (b) magnitude of $\hat{\tilde{U}}_1$.

In Figure 11, the nondimensional component $\hat{\tilde{U}}_1$ is reported versus the real and imaginary parts of the frequency $\tilde{s}$ for the wavevector $\tilde{\mathbf{k}}_0 = \begin{pmatrix} 1 & 0 \end{pmatrix}^T$, if the external forces are characterized by the forcing frequency $\tilde{S}_2 = 0.701\iota$ and unitary amplitudes. In this case the forcing frequency quasi-resonates with one of the metamaterial frequencies. Again, four poles in the forced response of displacement component $\hat{U}_1$ can be detected (Figure 11a). As expected, the pole associated to the associated to the forcing frequency (point $G_2$ marked by green dot) is found to lie in the closeness of a characteristic pole (point $A_2$ marked by yellow dot) of the material. The magnitude of the complex-valued component $\hat{U}_1$ is found to rapidly but continuously grow up in the closeness of all the poles (and their complex conjugates), approaching infinite values (Figure 11b). In correspondence of the two close poles, the quasi-resonance condition is found to determine the mutual interaction between a pair of undistinguishable peaks.

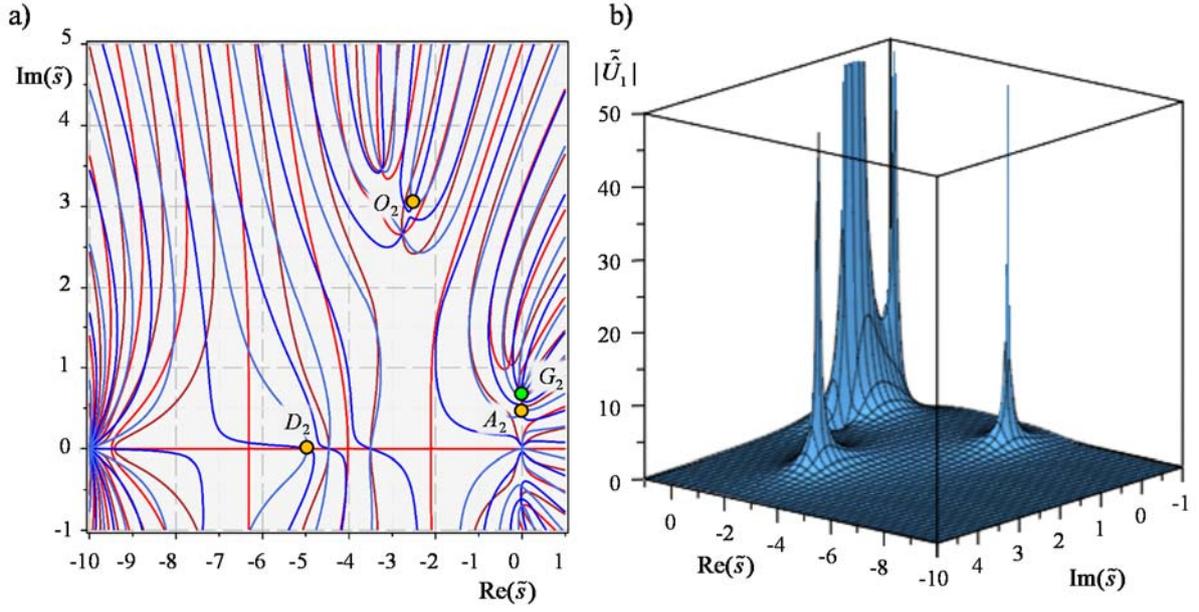

Figure 11. Response of the particular metamaterial $\mathcal{M}$ to quasi-resonant force ($\tilde{S}_2 = 0.701\imath$): (a) loci of the null real and null imaginary parts of the displacement component $\hat{U}_1$-numerator (curves in red-scale) and $\hat{U}_1$-denominator (curves in blue-scale), (b) magnitude of $\hat{U}_1$.

In order to reconstruct in the time-domain the damped response of the metamaterial to the external harmonic force with complex frequency $S$, the two anti-transformations (14) and (15) should be applied to determine the displacement vector $\mathbf{U}(S,t)$. Without loss of generality, the anti-transformations (15) can be applied to the integrand function $\hat{\mathbf{U}}(S_1,\mathbf{k}_1,s)$ of the anti-transformations (14) for a single selected value $\mathbf{k}_1$ of the wavevector, for the sake of simplicity. According to this idea, the real and imaginary parts of the first anti-transformed component $U_1(\mathbf{k}_1,t) = \mathcal{L}^{-1}\left[\hat{U}_1(S_1,\mathbf{k}_1,s)\right]$ of the vector $\hat{\mathbf{U}}(S_1,\mathbf{k}_1,s)$ can be analyzed. In Figure 12, the time-histories of the nondimensional real and imaginary parts $\mathrm{Re}\left(\tilde{U}_1(\tilde{\mathbf{k}}_1,\tau)\right)$ and $\mathrm{Im}\left(\tilde{U}_1(\tilde{\mathbf{k}}_1,\tau)\right)$ are reported in the nondimensional $\tau$-time domain for $\tilde{\mathbf{k}}_1 = (\pi\ 0)^T$. Two different time-histories are compared, corresponding to a harmonic non-decaying external force (with purely imaginary frequency $S_1 = 3\imath$) and a harmonically decaying external force (with complex frequency $S_1 = -1/100 + 3\imath$), respectively. The comparison shows that – after a short transient – the stationary damped response to the non-decaying external force (gray time histories) oscillates with constant amplitude at the frequency of the external force. On the contrary, the damped response to the decaying external force (red time histories) oscillates with exponentially decreasing amplitudes. No significant differences can be detected in the amplitudes of the real and imaginary parts of the complex valued response, whose respective phases are in quadrature, as expected. In Figure 13, the time-histories of the nondimensional real and imaginary parts $\mathrm{Re}(\tilde{U}_1(\tilde{\mathbf{k}}_1, \tau))$ and $\mathrm{Im}(\tilde{U}_1(\tilde{\mathbf{k}}_1, \tau))$ are reported for $\tilde{\mathbf{k}}_1 = (\pi/4\ 0)^T$. Again, two different time-histories are considered, corresponding to the harmonic non-decaying external force (with $S_1 = 3\imath$) and the harmonically decaying external force (with $S_1 = -1/100 + 3\imath$), respectively. The comparison shows that – after a significantly longer transient – the damped response to the non-decaying external force (gray time histories) tends to stationary oscillations with constant amplitude. The damped response to the decaying external force (red time histories) shows a long transient regime of oscillations, again with decreasing amplitudes.

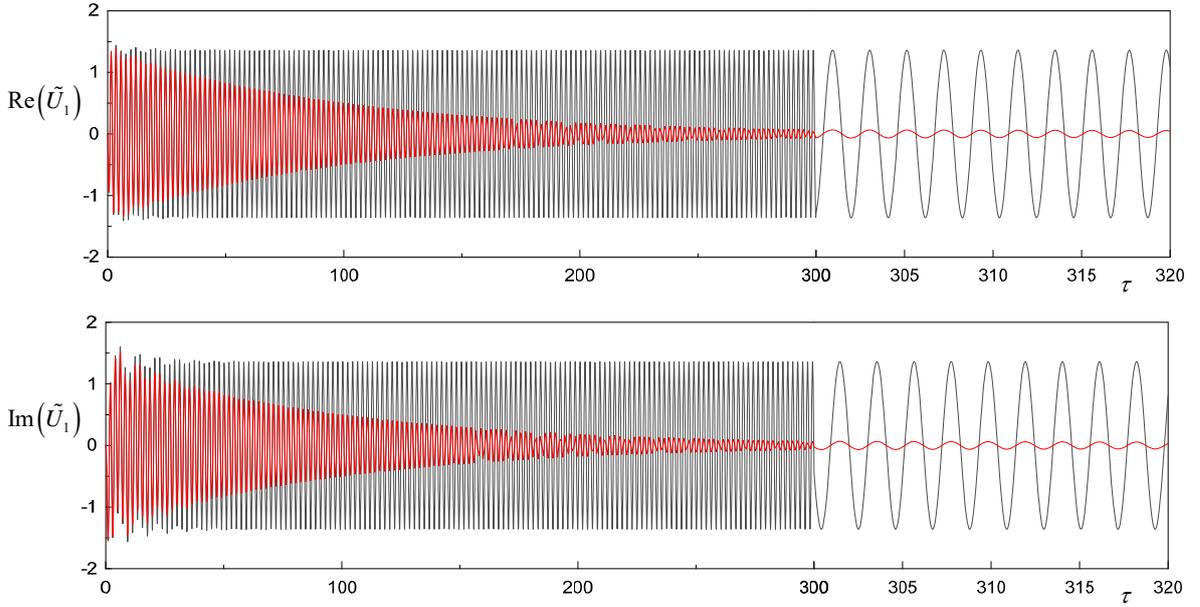

Figure 12. Time domain response of the particular metamaterial $\mathcal{M}$ to harmonically non-decaying external force ($\tilde{S}_1 = 3\imath$, gray curves) and harmonically decaying external force ($\tilde{S}_1 = -1/100 + 3\imath$, red curves) at fixed wavevector $\tilde{\mathbf{k}}_1 = (\pi\ 0)^T$: (a) nondimensional real part of $\tilde{U}_1(\tilde{\mathbf{k}}_1, \tau)$; (b) nondimensional imaginary part of $\tilde{U}_1(\tilde{\mathbf{k}}_1, \tau)$.

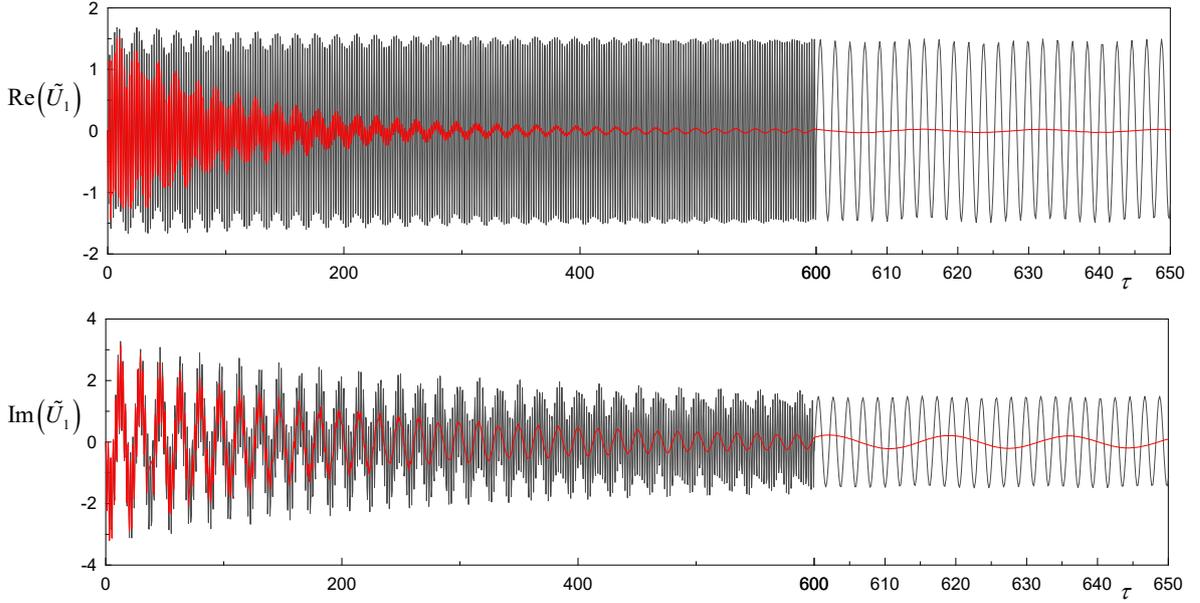

Figure 13. Time domain response of the particular metamaterial $\mathcal{M}$ to harmonically non-decaying external force ($\tilde{S}_1 = 3\iota$, gray curves) and harmonically decaying external force ($\tilde{S}_1 = -1/100 + 3\iota$, red curves) at fixed wavevector $\tilde{\mathbf{k}}_1 = (\pi/4 \ \ 0)^T$: (a) nondimensional real part of $\tilde{U}_1(\tilde{\mathbf{k}}_1, \tau)$; (b) nondimensional imaginary part of $\tilde{U}_1(\tilde{\mathbf{k}}_1, \tau)$.

## 6. Conclusions

A general mechanical formulation has been presented for describing the linear wave dynamics of beam lattice materials, characterized by a periodic cellular microstructure composed by a geometrically repetitive pattern of rigid rings interconnected by flexible ligaments. The non-dissipative microstructure of the beam lattice has been enriched by introducing auxiliary damped oscillators, housed by the periodic rings and purposely tuned to realize an acoustic metamaterial by exploiting the dynamic mechanism of local resonance. Each auxiliary oscillator, or resonator, has been viscoelastically coupled with the hosting ring. As peculiar aspect, the viscoelastic ring-resonator coupling has been derived by a proper mathematical formulation based on the Boltzmann superposition integral, whose kernel has been expressed by a Prony series. Accordingly, the free damped dynamics of the periodic cell is governed by a linear homogeneous system of integral-differential equations of motion. Therefore, imposing the quasi-periodicity conditions according to the Floquet-Bloch theory and applying the bilateral Laplace transform, a linear coupled system of ordinary differential equations with frequency-dependent coefficients has been ascertained to govern the free damped propagation of vibration waves. Consequently, the associated nonlinear, non-polynomial eigenproblem has been stated and numerically solved to determine the complex-valued dispersion spectrum of the viscoelastic metamaterial characterized by a quadrilateral periodic cell.

The acoustic and optical branches characterizing the complex-valued frequencies of the dispersion spectrum along the triangular boundary of the first Brillouin zone spanned by real-valued wavenumbers have been analyzed. Particularly, the complex spectra corresponding to different Taylor series approximations of the frequency-dependent rational coefficients governing the eigenproblem have been investigated. The classic eigenproblem and the complex spectrum associated to the standard dynamic equations with linear viscous damping have been recovered at the first-order approximation. Due to the non-polynomial nature of the eigenproblem coefficients, the exact eigensolution is characterized by a number of complex spectral branches that can exceed the model dimension. Exceeding spectral branches characterize also the eigensolution for high-order approximations of the eigenproblem coefficients. From a qualitative viewpoint, these exceeding branches have been found to enrich the purely real-valued part of the complex spectrum, corresponding to standing waves that do not propagate in space but are damped in time. Focusing on the propagating waves from a quantitative viewpoint, the exact and approximate eigensolutions shows that low-order approximations may determine non-negligible spectral effects, including the over-estimation or under-estimation of the stop bandwidth separating the acoustic from the optical branches.

Finally, the forced dynamics of the viscoelastic metamaterial under the effects of harmonically decaying and non-decaying waves of external mono-frequent forces acting on the microstructure has been investigated. The metamaterial response has been determined and parametrically analyzed in terms of dynamic compliance matrices and displacement components, both in the frequency and the time domains. In particular, the existence of poles and peaks dominating the complex frequency response functions has been discussed and interpreted on the light of the metamaterial spectrum, by recognizing the occurrence of resonant, quasi-resonant and non-resonant conditions.

## Acknowledgments

The authors acknowledge financial support of the Italian Ministry of Education, University and Research in the framework of the research MIUR Prin15 project 2015LYYXA8 "Multi-scale mechanical models for the design and optimization of micro-structured smart materials and metamaterials". The authors also acknowledge financial support by National Group of Mathematical Physics (GNFM-INdAM).

## Appendix A. Treatment of the eigenvalue problems

A mathematical technique can be sketched to attack the high-order polynomial eigenvalue problem governing the free damped wave propagation. The third order eigenvalue problem in the unknown eigenvalue $s$ is considered as benchmark, whereas higher order polynomial or rational problems can be reformulated and treated in a similar manner. In particular, the rational eigenvalue problem rising up from wave propagation problem under investigation (for the exact relation functions) can be reduced to a third order eigenvalue problem.

The third order eigenvalue problem $\mathbf{C}(\mathbf{k},s)\hat{\mathbf{U}} = \mathbf{0}$ can be expressed by separating the $s$-orders of the governing matrix $\mathbf{C}(\mathbf{k},s)$, yielding

$$\left(\mathbf{C}^{(0)} + s\mathbf{C}^{(1)} + s^2\mathbf{C}^{(2)} + s^3\mathbf{C}^{(3)}\right)\hat{\mathbf{U}} = \mathbf{0} \tag{A.1}$$

where the bracketed superscript of the matrices $\mathbf{C}^{(0)}, \mathbf{C}^{(1)}, \mathbf{C}^{(2)}$ and $\mathbf{C}^{(3)}$ denotes the $s$-order. Therefore, the polynomial eigenvalue problem (A.1) can be expressed in an equivalent linear form $\mathbf{L}(s)\mathbf{V} = \mathbf{0}$ where $\mathbf{V} = \begin{pmatrix} s^2\hat{\mathbf{U}} & s\hat{\mathbf{U}} & \hat{\mathbf{U}} \end{pmatrix}^T$ and the matrix $\mathbf{L}(s)$ can be written as

$$\mathbf{L}(s) = s\begin{bmatrix} \mathbf{C}^{(3)} & \mathbf{0} & \mathbf{0} \\ \mathbf{0} & \mathbf{I} & \mathbf{0} \\ \mathbf{0} & \mathbf{0} & \mathbf{I} \end{bmatrix} + \begin{bmatrix} \mathbf{C}^{(2)} & \mathbf{C}^{(1)} & \mathbf{C}^{(0)} \\ -\mathbf{I} & \mathbf{0} & \mathbf{0} \\ \mathbf{0} & -\mathbf{I} & \mathbf{0} \end{bmatrix} \tag{A.2}$$

Consequently, the eigenvalues solving the polynomial eigenvalue problem (A.1) coincide with the $s$-values that make the auxiliary higher-dimensional matrix $\mathbf{L}(s)$ singular. The corresponding eigenvectors coincide with a subvector of the vector $\mathbf{V}$.